\documentclass[referee] {aa}
\begin{document}

   \thesaurus{13
              (13.07.1;
               13.07.3;
                           )}

   \title{Comments on the paper ``Polarimetric Constraints on the
Optical Afterglow Emission from GRB 990123'' by Hjorth et al. (Science,  26
 1999)}


   \author{Abhas Mitra}


   \institute{Theoretical Physics Division,
   Bhabha Atomic Research Centre\\
Mumbai 400085, India,~ amitra@apsara.barc.ernet.in }

   \date{Received ; ~ Accepted }
\authorrunning{Mitra}
\titlerunning{ GRB polarization}
   \maketitle

\begin{abstract}
GRB 990123 is the most luminous event detected so far, and in an important
paper, Hjorth et al. (\cite{HJ}) reported an upper limit on the degree of
linear polarization of the optical afterglow for this burst ($P <2.3\%$).
One of the interprtations for this small value of $P$ was that the
emission was probably due a relativistic jet with ordered magnetic field,
and the viewing angle in the lab frame $\theta^\prime \la \Gamma^{-1}$,
where $\Gamma$ is the bulk Lorentz factor of the jet at the time of the
optical emission.  We point out that this conclusion resulted from a
confusion between the angles measured in the lab frame and in the plasma
rest frame. For ordered magnetic field, one would actually obtain a large
value of $P$ because the above mentioned angle would correspond to  a very
large angle, $\theta \la \pi/2$, in the plasma rest frame. And this is
probably the case with the blazars. On the other hand, it is indeed
possible to have $P\approx 0$ if the magnetic field of GRB 990123 was
completely chaotic and the viewing angle was considerably smaller than the
semi-angle of the jet.

 \end{abstract}

The above mentioned paper by Hjorth et al. (\cite{HJ}) is very important
in that it imposes an upper limit on the degree of linear polarization
($P< 2.3 \%$) for the optical afterglow of most luminous  Gamma Ray Burst
990123. When corrected for various effects which might affect the true
intrinsic polarization, this paper gave a value of $P=0.0\pm 1.4\%$, and
concluded that ``this is consistent with no linear polarization'', and an
upper limit of $P<2.3\%$ was set at 95\% confidence limit.

Hjorth et al. then attempted to interpret this result. Since a spherical
fireball too can develop a somewhat clumpy magnetic field structure and
may result in a significant value of $P<10\%$ (Gruzinov \& Waxman
\cite{GW}), Hjorth noted that the observed low value of $P$ is consistent
with emission from a spherical fireball. Then they also explored whether
such a low value of $P$ is consistent with emission from a narrow ultra
relativistic jet. Specifically, they considered the possibility whether
the observed value of $P$ could be very low if the jet is observed at an
angle $\theta^\prime\la \Gamma^{-1}$ where $\Gamma$ is the bulk Lorentz
factor of the jet at the time of optical emission. The reason that they
considered this particular value of $\theta^\prime$ is the following:  For
a random GRB event, ``for smaller viewing angles, the solid angle
decreases and for larger angles the flux drops''. At the time of the
optical emission, the estimated value of $\Gamma \sim 10-20$ and the
corresponding $\theta^\prime\sim 3^\circ -6^\circ$. To seek an answer to
the question mentioned above, Hjorth et al. banked on a work by Celloti \&
Matt (\cite{CM}, CM)
\cite{CM}.  From Fig. 2 \& 3 of CM, Hjorth et al. concluded that the value
of $P\approx 0$ for $\theta^\prime \sim 3^\circ -6^\circ$, :

``The measured value of polarization is therefore consistent with a small
angle between the jet axis and our line of sight''.

We would like to point out here that while arriving at this conclusion,
Hjorth et al. confused between the {\em angle measured in the observer's
frame} ($\theta^\prime$) and the angle measured {\em in the rest frame of
the plasma} emitting the radiation ($\theta$). The absissa of Figs. 2 and
3 in CM is $\theta$ and not $\theta^\prime$ contrary to what has been
considered by Hjorth et al. For small $\theta'$, these two angles are
connected by the well known special relativistic formula

\begin{equation}
\sin \theta = {2 \Gamma \theta^\prime \over 1 + \Gamma^2 {\theta^\prime}^2}
\end{equation}

Thus for $\theta^\prime \la \Gamma^{-1}$, $\theta \la \pi/2$! In other
words, in plasma rest frame, the small lab frame angle translates into a
very large value.  Then the Figs. 2 and 3 of CM would suggest a large
value of $P \sim 22\%$ rather than $P\approx 0\%$. In the most popular
classification scheme of the Active Galactic Nuclei, the viewing angle is
smallest for the radio selected blazars and they are indeed found to have
very high values of $P\sim 10-40\%$. In fact this point was clearly
mentioned by CM:

``As already mentioned, relativistic beaming effects, which are believed
to affect the emission of blazars, imply that the observed radiation can
be emitted at large angles in the plasma frame, even if the line of sight
is close to the jet axis. Therefore, these sources could also show high
X-ray polarization.''

Thus, the interpretation by Hjorth et al., in the framework of the work by
CM, that the absence of polarization is consistent with a jet
interpretation is incorrect and resulted from a confusion between angles
measured in the lab frame and plasma rest frame. For the so-called
galactic micro-quasars, it is believed that the viewing angle of the jet
is larger as compared to the radio selected blazars The micro-quasars too
display a fairly high value of $P\sim 10-15\%$ which is however, smaller
than the average linear polarization observed by the radio-selected
blazars (Mirabel \& Rodriguez \cite{Mira}). And thus these two objects
broadly support a scheme in which the degree of linear polarization
increases with decreasing viewing angle of the relativistic jet.
Physically this implies that a relativistic turbulent astrophysical jets
may be endowed with some ordered magnetic field.

Having said this, we emphasize that, if one moves away from the uniform
magnetic field configuration of CM, it might indeed be possible that
$P\approx 0$ for a very small $\theta^\prime $ for the highly unusual case
where there is no large scale ordered magnetic either along the axis of
the jet or in a direction perpendicular to it (in such a case, however,
it would be difficult to undrstand the phenomenon of jet confinement and
acceleration). In particular, if the magnetic field lies in a plane
perpendicular to the jet axis and is disordered on large scale, one would
obtain zero linear polarization if the jet is viewd along the jet axis.
This would be so because there would be total (axial) symmetry around the
viewing direction.  However, even in this case, if the jet is viewed
off-axis, there would be a temporally variable finite $P$  which can be
parameterized as (Ghisellini \& Lazzati \cite{GL})
\begin{equation}
P_{max} \approx 0.19 P_0 \left({\theta\over
\theta_c}\right)~\left(0.05\leq {\theta\over \theta_c} \le 1;~ 1^\circ \leq
\theta_c\leq 15^\circ \right)
\end{equation}
Here $P_0\sim 60-70\%$ is the intrinsic synchrotron polarization,
$\theta_c$ is the semi-angle of the jet, and a radiation spectral index of
0.6 is assumed. Also, now, we have dropped the prime from the lab frame angle.
Thus the value can reach
$\sim 10\%$ if the jet is viewed along its edge. 
This conclusion is reinforced by model of relativistic 
jet acceletation by supposed completely tangled magnetic field (Heinz \&
Begelman \cite{HB}).

However if supposed GRB jets are to be modelled on their AGN counterparts,
it may be necessary to consider the existence of some ordered magnetic
field in order to understand jet collimation away from the central engine,
and one may look forward to detect even much higher value of $P$.

Coming back to the work of CM, it dealt with self Snchrotron Compton model
of radiation emission, whereas GRBs afterglows are generally explained as
direct Snychrotron emission. For a given magnetic field configuration, the
polarizationn is in general higher in  the latter case.  To conclude,
although the result obtained by Hjorth et al. that $P  <2.3 \% $ for the
optical afterglow of GRB 990123 is important, the interpretation that this
low value of $P$ could be understood if the viewing angle is $\la
\Gamma^{-1}$ is incorrect if there is an ordered magnetic field in the
jet. On the other hand,  such a low value of $P$ may occur if the jet
magnetic field is completely tangled with  no component of field along the
axis, and if the viewing angle is much smaller than the semi-angle of the
jet.  If we accept this latter interpretation, we need to understand the
following: If there is usually an ordered magnetic field, over and above
small scale chaotic magnetic field, in relativistic turbulent plasma
associated with blazars  on scales of $\sim 100$pc or more and for the so
called micro-quasars,  why the relativistic plamsa responsible for the
emission of GRB 990123 was completely turbulent on all scales. Considering
all such aspects, it seems that the afterglow of GRB 990123 was more
likely to be quasispherical rather than jet type.  This is so because even
for a quasispherical afterglow, the observed spectral steepening can be
explained if it were propagating within a thick presupernova wind (Dai \&
Lu \cite{DL1}).  Yet, we caution that no definitive stand should be taken
at this juncture.  A broader discussion on the topic of jet collimation,
acceleration, magnetic field generation and expected degree of linear
polarization is beyond the scope of this research note.

\end{document}